\documentclass[twocolumn,showpacs,preprintnumbers,amsmath,amssymb]{revtex4}

\usepackage{epsfig,psfrag}
\usepackage{dcolumn}
\usepackage{bm}
\usepackage{graphicx}
\usepackage{color}

\begin{document}

\title{Analytical solution of the bosonic three-body problem}

\author{Alexander O. Gogolin,$^{1,2}$ Christophe Mora,$^3$ and 
Reinhold Egger$^2$}
\affiliation{
${}^{1}$ Department of Mathematics, Imperial College London, 180 Queen's Gate,
London SW7 2AZ, UK \\
${}^2$~Institut f\"ur Theoretische Physik, 
Heinrich-Heine-Universit\"at, D-40225 D\"usseldorf, Germany\\
${}^3$~Laboratoire Pierre Aigrain, ENS, 
Universit\'e Denis Diderot 7, CNRS; 24 rue Lhomond, 75005 Paris, France}

\date{\today}

\begin{abstract}
We revisit the problem of three identical bosons in free
space, which exhibits a universal hierarchy of bound 
states (Efimov trimers).  Modelling a narrow Feshbach resonance within a 
two-channel description, we map the integral equation for the 
three-body scattering amplitude to a one-dimensional 
Schr\"odinger-type single-particle equation, where 
an analytical solution of exponential accuracy is obtained.
We give exact results for the trimer binding energies, 
the three-body parameter, the 
threshold to the three-atom continuum, and the recombination rate.
\end{abstract}


\maketitle

The physics of ultracold atoms continues to generate immense
interest due to the exceptional tunability and perfection of the 
resulting many-body systems \cite{leggett}.
  A particularly interesting topic,
first raised in the context of nuclear physics, concerns three-body
bound states (trimers) of identical bosons.  
In seminal work done several decades ago, Efimov 
\cite{efimov1,efimov2,efimov3} showed that there exists a 
universal hierarchy of trimer states in the resonant case; 
for a recent review, see
Ref.~\cite{review}.
While these states were never observed in nuclear physics, first experimental
reports of Efimov states in the context of cold atoms have been 
recently published \cite{grimm}
(but see Ref.~\cite{esry} for a different interpretation of these results),
leading to renewed interest in Efimov physics also among theorists 
\cite{petrov,macek}. 
For instance, when atoms are confined in a trap, confinement effects 
can affect Efimov states in interesting ways \cite{jonsell,mora,castin}.
More generally, Efimov physics has strong experimental
relevance since it imposes limits on the atomic cloud lifetime 
via three-body recombination processes.

Here we reconsider the homogeneous three-dimensional (3D)
case for three identical bosons 
and present a novel solution of the Efimov problem, 
which is simpler than existing approaches 
\cite{review} and thus allows for analytical progress.
Our scheme is based on a mapping of the three-body integral 
equation to an effective 1D Schr\"odinger-type single particle 
equation, where quantum-mechanical intuition is available.
Zero-range interactions for three bosons is an ill-defined problem,
with a dense~\cite{danilov,minlos} and 
unbounded spectrum (`Thomas collapse')~\cite{thomas} of bound states.
In addition to the usual two-body scattering length $a$~\cite{leggett},
the regularization of the three-boson problem 
thus requires an additional parameter $R^*>0$ related to the 
finite range of the interaction potential.
We here consider the case of a narrow Feshbach resonance, 
where the short-distance regularization
does not involve higher angular momentum ($l>0$) partial waves \cite{petrov}. 
Remarkably, the resulting Schr\"odinger-type equation 
for the 1D motion of a fictitious quantum particle in a certain
potential, see Eq.~(\ref{stmfin}) below, can be solved analytically.
As is shown below, corrections to this solution  
are exponentially small at low binding energies. 
Our theory reproduces the known binding energy hierarchy
for shallow trimer states, see Eq.~\eqref{efimov}. 
We provide exact results 
for three key quantities: (i) the three-body parameter $\kappa_*$, 
see Eq.~\eqref{threebody}, which determines
many observables known to be universal functions of $\kappa_*$ 
\cite{review}; (ii) the scattering length $a_*^\prime$ 
where the Efimov states disappear into the
three-atom continuum, see Eq.~\eqref{threshold}; and  
(iii) the recombination rate
$\alpha_{\rm rec}$ for three-body recombination to
a weakly bound level, see Eq.~\eqref{recomb}.
Our result for  $\kappa_*$ depends on short-distance 
properties and is specific to narrow resonances, while
the $a_*^\prime$ and $\alpha_{\rm rec}$ results
apply for any type of resonance. 
Finally, our approach should also 
simplify future calculations in confined geometries. 

To begin with, let us derive a suitable three-body integral
equation encoding Efimov physics.
Inspired by Ref.~\cite{castin2},
we use a two-channel model to regularize the theory, 
where the boson operator $a_{\bf k}$, describing a 
momentum-${\bf k}$ atomic state, is coupled to the boson operator
 $b_{\bf K}$
describing a closed channel (molecular) state of momentum ${\bf K}$
and energy offset $E_0$. This model reproduces basic features
of Feshbach resonances~\cite{revk,castin3}, which are commonly used
to tune the interaction strength in experiments.
With coupling strength $\Lambda$ and $\epsilon_k= \hbar^2 k^2/2m$, 
the second-quantized many-particle Hamiltonian 
for a narrow Feshbach resonance is
\begin{eqnarray}\label{2ch}
H &=& \sum_{\bf k} \epsilon_k a_{\bf k}^\dagger
a_{\bf k}^{} + \sum_{\bf K} (E_0+ \epsilon_K /2 ) 
b_{\bf K}^\dagger b_{\bf K}^{} \\ \nonumber &+& \Lambda
\sum_{{\bf k},{\bf K}} \left( b^\dagger_{\bf K} a^{}_{{\bf k}+{\bf K}/2}
a_{-{\bf k}+{\bf K}/2}^{} + {\rm h.c.} \right).
\end{eqnarray}
The  two-body problem is solved by the {\sl Ansatz} 
 $|\psi\rangle = ( \beta b_{{\bf K}=0}^\dagger + \sum_{\bf k} A_{\bf k} 
a^\dagger_{\bf k} a^\dagger_{-{\bf k}} ) |0\rangle$,
where the Schr\"odinger equation $(H-E)|\psi\rangle=0$  implies
coupled equations for the complex coefficients $\beta$ and $A_{\bf k}$
\cite{castin3},
\begin{eqnarray}\label{2body}
(E_0-E) \beta +2\Lambda \sum_{\bf k} A_{\bf k} &=& 0 , \\  \nonumber 
\Lambda \beta + (2\epsilon_k-E) A_{\bf k} &=& 0. 
\end{eqnarray}
The second equation is solved by an incoming scattering state 
of momentum ${\bf k}_0$ and 
energy $E=2\epsilon_{k_0}= \hbar^2 k_0^2/m$,
where
$A_{\bf k}= (2 \pi)^3 \delta({\bf k}-{\bf k}_0) + \frac{4\pi f}
{k^2-k_0^2-i0^+},$
with the two-body scattering amplitude $f=- m\Lambda\beta/(4\pi\hbar^2)$.
We define
\begin{equation} \label{parameter}
\frac{1}{a} = \frac{1}{\ell} - \frac{2\pi\hbar^2 E_0}{m\Lambda^2 },
\quad R^* = \frac{2\pi \hbar^4}{m^2 \Lambda^2},
\end{equation}
where $\ell$ is a short-distance cutoff length \cite{footl}
of the order of the van der Waals potential size~\cite{castin2}. 
For a narrow Feshbach resonance, one has $R^*\gg \ell$ \cite{petrov,massignan}.
Equation (\ref{2body}) results in $f(k_0)= - [a^{-1} + ik_0 + 
R^* k_0^2]^{-1}$,
which coincides with the familiar effective range expansion \cite{leggett},
but is not restricted to the regime $|k_0 R^*|\ll 1$.
The three-body problem can then be solved using the {\sl Ansatz}
\[
\sum_{\bf K}\left( \beta_{\bf K}
b^\dagger_{\bf K} a^\dagger_{-{\bf K}} + \sum_{\bf k} A^{}_{{\bf K},{\bf k}} 
a_{{\bf k}+{\bf K}/2}^{\dagger} a_{-{\bf k}+{\bf K}/2}^{\dagger}
a_{-{\bf K}}^{\dagger} \right)|0\rangle,
\]
without the need for any additional regularization.
The (unsymmetrized) three-boson wavefunction 
$A_{{\bf K},{\bf k}}$ can be taken as even function of ${\bf k}$.  
The Schr\"odinger equation again leads to coupled
algebraic equations,
\begin{eqnarray} \label{3ch}
&& \left ( E_0 - E + \frac{3}{4} \frac{\hbar^2 K^2}{m} 
\right) \beta_{\bf K} \\ \nonumber
&& + 2\Lambda \sum_{\bf k} \left( A_{{\bf K},{\bf k}} 
+2A_{{\bf k}-{\bf K}/2,-{\bf k}/2-3{\bf K}/4} \right) = 0,
\\ \nonumber && \Lambda \beta_{\bf K} +  \left(2\epsilon_k - E  + \frac{3}{4}
\frac{\hbar^2 K^2}{m} \right) A_{{\bf K},{\bf k}}=0.
\end{eqnarray}
Since we search for bound states, we now put $E=-\hbar^2 \lambda^2 /m<0$.
The second equation is then solved by 
$A_{{\bf K},{\bf k}}= - (m\Lambda \beta_{\bf K}/\hbar^2) 
[3 K^2/4 + k^2 + \lambda^2 ]^{-1}$, while the first determines the 
three-body scattering amplitude $\beta_{\bf K}$.
Employing the definitions of $a$ and $R^*$ in Eq.~(\ref{parameter}), we find
\begin{equation}\label{int1}
(3 R^* K^2 /4 + \hat{L}_{\lambda} - 1/a ) \beta_{\bf K} =0,
\end{equation}
where we introduce the operator
\begin{eqnarray}\label{L0}
\hat{L}_{\lambda} \beta_{\bf K} &\equiv &
\left( \sqrt{\lambda^2+3K^2/4} + 
\lambda^2 R^* \right) \beta_{\bf K} \\ \nonumber
& -& \frac{1}{\pi^2} \int d^3 K' \frac{\beta_{{\bf K}'}}{ K^{\prime 2} +K^2+{\bf K}'\cdot {\bf K}+\lambda^2}.
\end{eqnarray}
This is the three-boson integral equation, cp.~Refs.~\cite{stm,minlos,petrov}. 
In contrast with 
the commonly used Bethe-Peierls boundary
conditions or pseudo-potentials \cite{stm}, 
the Hamiltonian~\eqref{2ch} is manifestly self-adjoint,
and thus our two-channel description provides a natural regularization
scheme free of the mathematical difficulties 
encountered otherwise \cite{minlos,danilov}.
It differs from the real-space scheme used
originally by Efimov \cite{efimov1,efimov2,efimov3}, but
leads to the same universal trimer energy hierarchy.
Rotational symmetry expresses the solution of Eq.~(\ref{int1})
in terms of a `wavefunction' $\psi(k)$ for a fictitious  1D problem, 
$\beta_{\bf K} \equiv \psi(k)/k$, with $k\equiv |{\bf K}|>0$.
We thus integrate over the angular variables in
 Eq.~(\ref{L0}), and obtain
\begin{eqnarray} \label{stmeq}
&& \left( \sqrt{3 k^2/4 + \lambda^2 } - a^{-1} + R^*(3 k^2/4 + \lambda^2) 
\right) \psi(k)  
\\ \nonumber &&- \frac{2}{\pi}\int_0^\infty dk' \ln\left(
\frac{ k^2+kk'+k'^2+\lambda^2 }{ k^2-kk'+k'^2+\lambda^2}\right) 
\psi(k') = 0.
\end{eqnarray}
Since the logarithmic term is odd under $k'\to -k'$, we extend $k$ to 
include negative values, but require  $\psi(-k)=-\psi(k)$.  The 
integral in Eq.~(\ref{stmeq}) is then taken over all $k'$,
with the replacement $2/\pi \to 1/\pi$. 

We next show how Eq.~(\ref{stmeq}) can be mapped to 
a quantum-mechanical single-particle equation.
To that end, we employ the transformation
\begin{equation}\label{xidef}
k=\frac{2\lambda}{\sqrt{3}}\sinh\xi, \quad \xi \in \mathbb{R},
\end{equation} 
and search for antisymmetric solutions $\psi(\xi)=-\psi(-\xi)$.
Substitution (\ref{xidef}) is helpful because (i) 
the square root in Eq.~(\ref{stmeq}) rationalizes,
and (ii) the  logarithmic kernel becomes homogeneous.  
To see that, we decompose the logarithmic factor in Eq.~(\ref{stmeq}) as
$ \ln \left(\frac{e^{2\xi}+e^{\xi+\xi'}+e^{2\xi'}}{
e^{2\xi}-e^{\xi+\xi'}+e^{2\xi'}}\right) +
 \ln\left(\frac {e^{2(\xi+\xi')}-e^{\xi+\xi'}+1 } 
{e^{2(\xi+\xi')}+e^{\xi+\xi'}+1 } \right).$
Both terms yield the same contribution in Eq.~(\ref{stmeq}),
which is seen by
letting $\xi'\to -\xi'$ in the integral associated with the second
term and exploiting antisymmetry, $\psi(-\xi')=-\psi(\xi')$.
This leads us to the  difference kernel
$ T(\xi) = \frac{4\pi}{3\sqrt{3}} \delta(\xi) 
- \frac{4}{\pi \sqrt{3}} \ln \left( 
\frac{e^{2\xi}+e^\xi +1}{e^{2\xi}-e^{\xi}+1}\right),$
with Fourier transform
\begin{equation}\label{fourkern}
\hat T(s)= \int_{-\infty}^\infty d\xi e^{i s\xi} T(\xi)=
\frac{4\pi}{3\sqrt{3}} -
\frac{8}{\sqrt{3}} \frac{\sinh(\pi s/6)}{s\cosh(\pi s/2)}.
\end{equation}
This kernel acts on a test function $g(\xi)$ as a differential operator,
$\int_{-\infty}^\infty d\xi' T(\xi-\xi') g(\xi') = \hat T
\left(- i \frac{d}{d\xi} \right) g(\xi).$
The function (\ref{fourkern}) thus plays the role of a  {\sl kinetic
energy operator}.  For the standard non-relativistic Schr\"odinger equation, $\hat T(p)=p^2/2m$;
 our notation $s$ for `momentum' follows  convention \cite{review}. 
Here the dispersion relation~\eqref{fourkern}  starts as
$\hat T(s)\sim s^2$ at small momentum
and levels off to $4\pi/(3\sqrt{3})$ as $s\to \infty$.   It
is thus bounded from below and from above,
similar to what happens for the typical band structure of a solid.
Note that we slightly abuse terminology here,
since $\xi$ parametrizes the physical momentum, see Eq.~(\ref{xidef}),
and hence $s$ in reality corresponds to a spatial variable.  
Nevertheless, this analogy is quite helpful
to make contact with elementary textbook quantum mechanics,
and we will henceforth denote $s$ as `momentum'
conjugate to the 1D `space' variable $\xi$.
We also introduce the symmetric single-particle potential
\begin{equation}\label{potential}
U(\xi) = -\frac{1}{a\lambda} \frac{1}{\cosh\xi} + R^*\lambda \cosh \xi,
\end{equation}
and the `energy' ${\cal E} = 4\pi/(3\sqrt{3})-1 \simeq 1.41839$.
After rescaling $\psi(\xi)\to \psi(\xi)/\cosh(\xi)$, Eq.~(\ref{stmeq}) then
assumes the equivalent final form
\begin{equation}\label{stmfin}
\left[ \hat T\left(-i\frac{d}{d\xi}\right) + 
U(\xi)-{\cal E }\right]\psi(\xi)=0,
\end{equation}
subject to the antisymmetry condition $\psi(\xi)=-\psi(-\xi)$.
This equation formally describes the 1D motion of a quantum particle 
with non-standard dispersion relation (\ref{fourkern})
in the potential $U(\xi)$ at energy ${\cal E}$.

Let us now consider the {\sl resonant limit} $1/a=0$.
The solution of Eq.~(\ref{stmfin}) for $R^*=0$ 
is given by $\psi(\xi)\sim \sin(s_0\xi)$,
where $s_0\simeq 1.00624$ is the real and positive root of
$\hat T(s_0) =  {\cal E}$.
However, this does not lead to well-defined energy eigenstates 
\cite{review,thomas,danilov}.
We thus need to analyze the regularization mechanism due to $R^*> 0$,
which is quite simple to understand within our picture:
Potential (\ref{potential}) approaches $+\infty$ at
$\xi\to \pm\infty$, and hence all eigenstates must be quantized bound state 
solutions, similar to what happens for a simple 
harmonic oscillator.  In our case, the `energy' ${\cal E}$ is always fixed 
 but the true spectral parameter 
is $\lambda$: only those values of $\lambda$ are allowed 
(possibly a finite set or countable infinity), where a 
bound state with energy ${\cal E}$ exists.  These discrete values 
$\lambda_n$ (integer $n$) then determine the Efimov trimer bound state
energies $E_n= - \hbar^2\lambda^2_n/m$.  
For  $R^*\lambda\ll 1$,  we have $n \gg 1$, and zero energy 
represents a spectral accumulation point, see Eq.~\eqref{efimov} below.
Taking $\xi>0$ with $\psi(-\xi)=-\psi(\xi)$, 
the potential $U(\xi)$ can be neglected 
against ${\cal E}$ in the region
$\xi \ll \xi_*$, where $\xi_*= \ln [ 2/(R^*\lambda) ] \gg 1$.
Here the solution must therefore be
$\psi_1(\xi)=c_1\sin(s_0\xi)$, with complex amplitude $c_1$.
On the other hand, for all $\xi\gg 1$  (which includes
$\xi\approx\xi_*$), the potential takes the form 
 $U(\xi)=e^{\xi-\xi_*}$.  Shifting $\xi$ by $\xi_*$, this regime 
is thus described by  a {\sl universal}\ (parameter-free) equation
\begin{equation}\label{univ}
\left[ \hat T\left(-i\frac{d}{d\xi}\right) + 
e^\xi -{\cal E }\right]\psi(\xi)=0.
\end{equation}
Note that antisymmetry of $\psi(\xi)$ should not be imposed here
because of the shift. 
For $\xi\to -\infty$, the exponential term in this equation 
vanishes, and
the asymptotic behavior $\psi(\xi) \sim \sin(s_0\xi+\pi \gamma)$ with a
non-trivial phase shift $\gamma$ is expected.
Coming back to the original $\xi$,  the solution
for $1 \ll \xi \ll \xi_*$ is
$\psi_2(\xi)=c_2\sin[s_0 (\xi-\xi_*) + \pi \gamma]$,
where $c_2$ is another amplitude. Evidently, both $\psi_2$
and $\psi_1$ should match within the broad region $1\ll \xi \ll \xi_*$.
This implies the {\sl quantization condition}
\begin{equation}\label{quant}
\xi_*(\lambda_n) = \ln[2/(R^*\lambda_n)] = \pi(n+\gamma)/s_0,
\end{equation}
yielding the on-resonance {\sl Efimov trimer energies} 
\begin{equation}\label{efimov}
E_n =  -\frac{\hbar^2 \kappa_*^2 }{m } e^{-2\pi n/s_0}
\end{equation}
with the famous universal ratio $E_{n+1}/E_n=
e^{-2\pi/s_0} \simeq 1/515.03$ between subsequent levels.
Note that $R^*\lambda \sim e^{-\pi n/s_0}$ is exponentially
small for shallow Efimov trimers. Since all corrections to our 
approach are of order $R^*\lambda$,  
deviations from our results (e.g., $|(\delta E_n)/E_n|$) 
 are exponentially small for $n\to \infty$. 
The {\sl three-body parameter} $\kappa_*$ \cite{review},
defined only up to multiplicative factors of $e^{\pi/s_0}$,
is here taken as $\kappa_* R^*  = 2 e^{-\pi\gamma/s_0}$.
To obtain it, we now calculate $\gamma$ from the 
universal problem (\ref{univ}). 

Remarkably, the exact solution of Eq.~(\ref{univ}) 
can be obtained as a Barnes-type contour integral~\cite{whittaker,footnew},
\begin{equation}\label{barnes}
\psi (\xi) = \int_{-i\infty+0^+}^{i\infty+0^+} \frac{d\nu}{2\pi i} e^{-\nu\xi}
\,  C(\nu), 
\end{equation}
where Eq.~(\ref{univ}) implies the recurrence relation
$[\hat{T} (i\nu) - {\cal E}] C(\nu) = - C(\nu+1).$
Applying $\hat{T} -{\cal E}$ to Eq.~\eqref{barnes} thus changes the
integrand to  $- e^{-\nu\xi} \, C(\nu+1)$. The desired result 
$- e^\xi \psi(\xi)$ is obtained if no pole is crossed
when the contour is shifted back, $\nu \to \nu-1$, see also Ref.~\cite{macek}.  
Therefore $C(\nu)$ must have no
poles within the strip $0 <  {\rm Re} (\nu) \le 1$ of the complex plane.
Using the Weierstrass theorem~\cite{whittaker}, we can now express 
$ \hat T(i\nu)
-{\cal E} =  \prod_{p=0}^{\infty} \frac{\nu^2-u_p^2}{\nu^2-b_p^2} $
as a convergent infinite product 
in terms of its poles $\pm  b_p$, where $b_p\equiv 2 p +1$, and its 
zeroes $\pm  u_p$.
Two zeroes are on the imaginary axis, $u_0 = i s_0$, while all others are 
real and correspond to $u_1 = 4, u_2= 4.6, \ldots$~\cite{efimov2,review}.
The function satisfying the recurrence relation and the required 
analytic properties is 
$C(\nu)= [\pi / \sin( \pi (\nu-i s_0))] \, C_+ (\nu)$, where
\begin{equation}\label{defC}
C_{+} (\nu) = \prod_{p=0}^\infty \frac{\Gamma(\nu+u_p)\Gamma(1-\nu+b_p)}{
\Gamma(\nu+b_p) \Gamma(1-\nu+u_p)}
\end{equation}
with the gamma function $\Gamma(z)$.
In fact, $C(\nu)$ has no poles in the strip 
$0<{\rm Re}(\nu)<2$, while the poles 
at $\nu=2$ and $\nu=\pm is_0$ imply from
Eq.~\eqref{barnes} the asymptotic behaviors $\psi(\xi)\sim
e^{-2 \xi}$ for $\xi\to \infty$
 and, as expected, $\psi(\xi)\sim\sin(s_0\xi+\pi \gamma)$  for $\xi\to
-\infty$.
The phase $\gamma$ now follows from the ratio of the residues at the two
poles $\nu = \pm i s_0$. We find the exact result
\begin{equation}\label{gamma}
\gamma = \frac{1}{2}  -  \frac{1}{\pi} {\rm Arg} C_+ (i s_0)
 \simeq   - 0.090518155.
\end{equation}
Hence the three-body parameter is given by
\begin{equation}\label{threebody}
\kappa_* R^* = 2 e^{-\pi\gamma/s_0} \simeq 2.6531.
\end{equation}
This exact result roughly agrees  
with the available numerical estimate $\kappa_* R^*\approx 2.5$ 
\cite{petrov,review}. We have also performed
a more accurate numerical check \cite{footnew2}. This
gives $\kappa_* R^*\simeq 2.653(1)$, 
in good agreement with Eq.~\eqref{threebody}.

Using Eq.~(\ref{gamma}) we can also compute
the scattering length $a_*^\prime<0$ where the 
Efimov state joins the three-atom threshold at $\lambda\to 0$.
In that limit, the off-resonant potential (\ref{potential}) 
can be written as $U(\xi)=(2/|a|\lambda)e^{-\xi} + (R^*\lambda/2) e^\xi$, 
and the same phase shift $\gamma$ thus determines the behavior at 
$\xi\to \pm \infty$. 
Equation (\ref{quant}) then implies 
$\ln(2/R^*\lambda)=\pi(n+\gamma)/s_0$ and
$\ln(2/|a|\lambda) = \pi(n'-\gamma)/s_0$
(with integers $n,n'$).
This in turn gives $a_*^\prime /R^*=-e^{2\pi \gamma/s_0}$ (modulo factors 
of $e^{\pi/s_0}$), and hence the exact result for the three-atom
continuum threshold,
\begin{equation}\label{threshold}
a_*^\prime \kappa_* = -2e^{\pi \gamma/s_0} \simeq  -1.50763 ,
\end{equation}
consistent with the previously obtained 
numerical result $a_*^\prime \kappa_*=-1.56(5)$ \cite{review}.

Finally, let us briefly address the three-body recombination
rate $\alpha_{\rm rec}$.  For energy $E=0^+$ and $a>0$,
the solution of Eq.~\eqref{3ch} proceeds as before, but now
includes a source term in Eq.~\eqref{int1} \cite{castin}. Fourier transformation to
real space and
rescaling ${\bf r} \to \sqrt{3} {\bf r}/2$ as well as
$\beta ({\bf r}) \to 6 \sqrt{2 \pi R^*} a (1 + 4 \beta({\bf r})/\sqrt{3})$
results in 
\begin{equation}\label{diff1}
\left( - R^* \Delta_{\bf r} + \frac{2}{\sqrt{3}} \hat{L}_0 -\frac{1}{a} \right)
\beta ({\bf r}) = \frac{1}{r},
\end{equation}
where $\hat{L}_0$ is given in Eq.~\eqref{L0}.
For $r\to \infty$, the solution of Eq.~\eqref{diff1} contains
an outgoing spherical wave $A e^{i k_d r}/r$, describing  atom and dimer
separating after recombination, where 
$\hbar^2 k_d^2/m$ is the weakly bound dimer
energy calculated from $f^{-1}( i k_d)=0$.
Following Ref.~\cite{castin2},
 $\alpha_{\rm rec}$ is then obtained in the form
$\alpha_{\rm rec} =96 \sqrt{3} \pi^2 (\hbar a^2 /m) 
|A|^2 (1+2 R^* k_d).$ For $R^* \gg a$, 
$\hat{L}_0$ becomes negligible in Eq.~\eqref{diff1}, and the result
$\alpha_{\rm rec} =192 \sqrt{3} \pi^2 (\hbar/m)
\sqrt{a^7 R^*}$ \cite{petrov} follows.
For $R^* \ll a$, on the other hand, a complete analytical solution 
is also possible by relating Eq.~\eqref{diff1}
to the universal problem \eqref{univ}.
In fact, in the broad region $R^*\ll r\ll a$, Eq.~\eqref{diff1}
is basically the Fourier transformed equation (\ref{univ}), 
leading to $\beta (r) \sim  r^{-1} \sin [ s_0 \ln ( r/R_0) ]$,
with the renormalization length \cite{review}
$R_0 = R^* e^{ [\pi(\gamma+ 1/2) +  
{\rm Arg} \Gamma(i s_0)]/ s_0} \simeq 0.55890 R^*$.
This imposes a boundary condition 
for $R^* \ll r \ll a$ on solutions to Eq.~(\ref{diff1}), which 
in turn allows to compute $A$ and hence $\alpha_{\rm rec}$ in closed form.
Postponing details to elsewhere, the final result is, 
cf.~Refs.~\cite{macek,gasaneo},
\begin{equation}\label{recomb}
\alpha_{\rm rec} = \frac{2^7 \pi^2  (4 \pi -3 \sqrt{3})}{\sinh^2(\pi s_0)} \
\frac{\hbar a^4}{m} \sin^2 \delta_r
\end{equation}
with $\delta_r=\delta_1-{\rm Arg}(1+e^{-2\pi s_0} e^{2i\delta_1})$,
where $\delta_1=s_0 \ln(a\kappa_*/2)+\pi/2-\pi\gamma$.

To conclude, we have presented a fresh theoretical approach to the 
Efimov problem of three identical bosons. It is based on 
mapping the three-body integral equation for the scattering amplitude
to a  Schr\"odinger-like equation for a quantum-mechanical 
particle in a 1D potential.  For low binding energies, this permits
a solution of exponential accuracy, and we have derived exact results
for several key observables.   

We thank Y. Castin and F. Werner for discussions.
This work was supported by the SFB TR 12 of the DFG and by the
ESF program INSTANS. A.O.G. thanks the Humboldt foundation
for a Friedrich-Wilhelm-Bessel award enabling his extended stay
in D\"usseldorf.


\begin{thebibliography}{99}
\bibitem{footnew}
The exact solution can be fully justified in the framework of 
the theory of generalized hypergeometric functions.
\bibitem{footnew2}
The Efimov trimer disappears into the atom-dimer continuum
at the scattering length $a_*>0$ given by 
$a_*\kappa_* \simeq 0.070764$ \cite{review}. By numerical
solution of Eq.~(\ref{int1}), we have determined
$R^*/a_*$ and hence $R^*\kappa_*$.
\end{thebibliography}
\end{document}